\documentclass[twocolumn,showpacs,preprintnumbers,amsmath,amssymb,groupedaddress,prl]{revtex4}

\usepackage{amssymb}
\usepackage{amsfonts}
\usepackage{amsmath}
\usepackage{graphicx}
\usepackage{textcomp}
\usepackage{mathptmx}
\usepackage{dcolumn}
\usepackage{bm}
\usepackage{times}

\newcommand{\ket}[1]{\ensuremath{\left|#1\right\rangle}}

\begin{document}

\title{Photon Shot Noise Dephasing in the Strong-Dispersive Limit of Circuit QED}

\author{A. P. Sears}
\author{A. Petrenko}
\author{G. Catelani}
\author{L. Sun}
\author{Hanhee Paik}
\author{G. Kirchmair}
\author{L. Frunzio}
\author{L. I. Glazman}
\author{S. M. Girvin}
\author{R. J. Schoelkopf}

\affiliation{Department of Physics and Applied Physics, Yale
University, New Haven, Connecticut 06520, USA}
\date{\today }

\begin{abstract}
We study the photon shot noise dephasing of a superconducting
transmon qubit in the strong-dispersive limit, due to the coupling
of the qubit to its readout cavity. As each random arrival or
departure of a photon is expected to completely dephase the qubit,
we can control the rate at which the qubit experiences dephasing
events by varying \textit{in situ} the cavity mode population and
decay rate.  This allows us to verify a pure dephasing mechanism
that matches theoretical predictions, and in fact explains the
increased dephasing seen in recent transmon experiments as a
function of cryostat temperature. We investigate photon dynamics
in this limit and observe large increases in coherence times as
the cavity is decoupled from the environment. Our experiments
suggest that the intrinsic coherence of small Josephson junctions,
when corrected with a single Hahn echo, is greater than several
hundred microseconds.
\end{abstract}

\pacs{03.67.Lx, 42.50.Pq, 85.25}

\maketitle

Solid-state superconducting quantum systems offer convenient and
powerful platforms for quantum information processing.  Rapid
progress~\cite{houck_controlling_2008, neeley_transformed_2008,
steffen_decoherence_2009} is being made in engineering qubits and
effectively isolating them from the surrounding electromagnetic
environment. Despite these efforts, the measurement apparatus will
always be used to contact the environment and is therefore a
potential source for decoherence.

Recently~\cite{paik_observation_2011} superconducting qubits have
been created inside a three-dimensional (3D) resonator, leading to
more than an order of magnitude increase in coherence time.
Interestingly, the energy relaxation time $T_1$ has increased even
more than the phase coherence time $T_2^*$, pointing to a new or
newly important mechanism for dephasing~\cite{houck_life_2009}.
These devices have a single Josephson junction, eliminating the
sensitivity to flux noise~\cite{wellstood_low-frequency_1987}, and
surprisingly show only a weak temperature-dependent dephasing,
inconsistent with some predictions based on extrapolations of
junction critical current
noise~\cite{van_harlingen_decoherence_2004,eroms_low-frequency_2006}.
In these devices, the qubit state is detected by observing the
dispersive frequency shift of a resonant cavity. However, it is
known \cite{bertet_dephasing_2005, gambetta_qubit-photon_2006,
serban_crossover_2007} that in the strong-dispersive regime the
qubit becomes very sensitive to stray cavity photons, which cause
dephasing due to their random ac-Stark
shift~\cite{schuster_ac_2005}. It requires increasing care to
prevent this extrinsic mechanism from becoming the dominant source
of dephasing as qubit lifetimes increase.   Experiments
elsewhere~\cite{rigetti_superconducting_2012} and in our
lab~\cite{luyanandrei} have shown that pure dephasing times
can be many hundreds of microseconds with careful thermalization and more extensive filtering. 

In this Letter, we quantitatively test the dephasing of a qubit
due to photon shot noise in the strong-dispersive coupling limit
with a cavity.  In this novel regime where the ac-Stark shift per
photon is many times greater than the qubit linewidth $\gamma$ and
the cavity decay rate $\kappa$~\cite{Schuster_resolving_2007}, the
passage of any photon through the cavity performs a complete and
unintended measurement of the qubit state.  This limit also allows
a precise determination of the photon number in the cavity using
Rabi experiments on the photon number-split qubit
spectrum~\cite{johnson_quantum_2010}. With a simulated thermal
bath injecting photons into the cavity and $\textit{in situ}$
mechanical adjustment of the cavity $\kappa$, we find a pure
dephasing of the qubit that quantitatively matches
theory~\cite{gambetta_qubit-photon_2006}.  Furthermore, we verify
that the qubit is strongly coupled to photons in several cavity
modes and find that the dephasing from these modes accounts for
the reduced coherence times as a function of cryostat temperature.
Our measurements at 10~mK demonstrate that decreasing $\kappa$
leads to longer qubit coherence times, suggesting that existing
dephasing in superconducting qubits is due to unintended and
preventable measurement by excess photons in higher frequency
modes.

\begin{figure}
\includegraphics[width=3.375in]{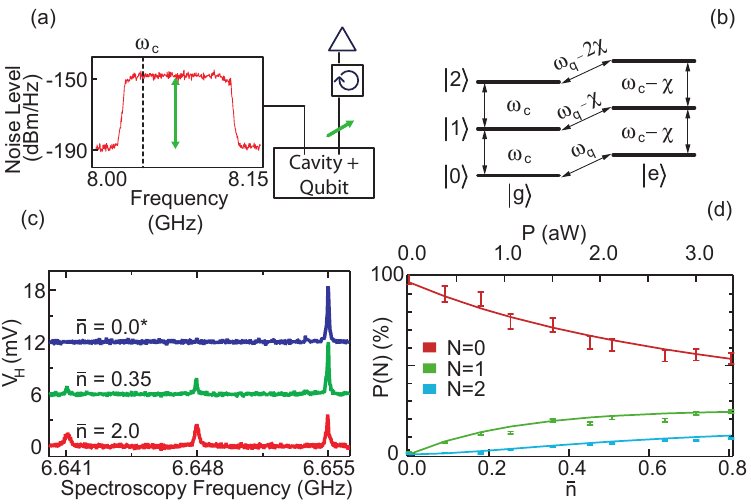} 
\caption{\label{rabiplot} (a) Experimental setup. Noise of varying
amplitude at the cavity transition frequency is sent into the
input port of a 3D resonator.  The output port of the resonator
has a movable coupler which varies the output side coupling
quality factor $Q_c$ from $1.0\times10^5$ to $2.5\times 10^7$. (b)
Energy level diagram. The qubit has a transition frequency that is
ac-Stark shifted by $-\chi$ for each photon in the cavity. (c)
Photon number-splitting of the qubit spectrum. We inject noise and
create a mean number $\bar{n}$ of photons in the fundamental mode.
The peaks correspond to $N = $ 0,1,2 photons from right to left,
with the cavity $Q=1\times10^6$, and (*) even without applied
noise we measure a photon occupation in the $\text{TE}_{101}$ mode
of the cavity to be $\bar{n}\sim 0.02$. (d) Cavity population.
Rabi experiments performed on each photon peak $N$ for increasing
noise power with cavity $Q=2.5\times10^5$. The signal amplitude
gives the probability of finding $N$ photons in the cavity.  Two
linear scaling factors, fit globally, provide conversion from
homodyne readout voltage~\cite{reed_high-fidelity_2010} to
probability (vertical axis), and from attowatts within the cavity
bandwidth to $\bar{n}$ (horizontal axis).  Error bars represent
$1\sigma$ fluctuations in the $\ket{e}$ state readout voltage. The
solid lines are a thermal distribution using the fit scaling
parameters.}\label{rabiplot1}
\end{figure}

The experiments were performed (see Fig. \ref{rabiplot1}) with a
transmon qubit coupled in the strong-dispersive limit to a 3D
cavity, and well approximated by the
Hamiltonian~\cite{nigg_BBQ_2012}:

\begin{equation}
 \text{H}_\text{eff}/\hbar= \omega_c
 a^{\dag}a+\left(\omega_q-\chi a^{\dag}a\right)b^\dag b - \frac{\alpha}{2} b^\dag b^\dag b
 b,
\end{equation}
where the operator $a^\dag$ creates a cavity photon and the
operator $b^\dag$ creates a qubit excitation.  Then $\omega_c$ is
the cavity frequency, $\omega_q$ and $\alpha$ are the qubit
frequency and anharmonicity, and $\chi/2\pi=7$~MHz is the light
shift per photon which can be 1000 times larger than the qubit
linewidth of $\gamma/2\pi=5-12$~kHz, and the cavity linewidth
$\kappa/2\pi=6-120$~kHz. The large dispersive shift leads to the
well-resolved peaks in the qubit spectrum shown in Fig.~1c,
allowing us to conditionally manipulate the qubit depending on the
cavity photon number $N$~\cite{johnson_quantum_2010}. Measuring
the height of a given photon number-split qubit peak (or the
amplitude of a Rabi oscillation at frequency $\omega_q-N\chi$)
allows a direct determination of the probability $P(N)$ for the
cavity to have a particular photon number.




Dephasing of the qubit can be caused by a random change in cavity
photon number, which shifts the qubit energy by $\hbar\chi$ per
photon and leads to a large rate of phase accumulation relative to
$\gamma$.  Then the pure dephasing rate $\gamma_\phi$, obtained in
a Ramsey experiment for the qubit, depends on the stability of the
$N$ photon cavity state. When the cavity is connected to a thermal
bath, the probability $P(N)$ follows a system of
equations~\cite{walls_quantum_1994} for the rate of change into
and out of the $N$ photon state: $dP(N)/dt=\kappa (\bar{n}+1)
(N+1) P(N+1) + \kappa\bar{n} N P(N-1) - \Gamma_\text{out}P(N)$,
where the cavity decay rate $\kappa = 1/\tau$ is the inverse of
its decay time $\tau$, $\bar{n}$ is the average number of photons,
and
\begin{equation}\label{gammaout}
\Gamma_{\text{out}} = \kappa \left[ (\bar{n}+1) N + \bar{n}(N+1)
\right]
\end{equation}
combines the spontaneous emission of photons with the stimulated
emission due to thermal photons.  Then, in the strong-dispersive
regime (and neglecting other sources of dephasing) the dephasing
rate becomes $\gamma_\phi=\Gamma_\text{out}$, and the success of
an experiment that relies on phase predictability of the qubit
requires a constant photon number in the cavity throughout each
cycle.

To verify this prediction for $\gamma_\phi$ quantitatively, we
first calibrate our thermal bath and then obtain $\kappa$ with
experiments on the photon peaks of the qubit.  We can determine
the cavity decay rate $\kappa$ by exciting the cavity with a short
coherent pulse while measuring the repopulation of the ground
state $\ket{\text{g,0}}$ (i.e. the amplitude of the zero-photon
Rabi oscillations) over timescale $\tau$.  Alternatively, exciting
the cavity with a wideband noise source that covers the cavity
$\omega_c$ transition frequency but not the qubit $\omega_q$
transition frequency, creates an average photon number $\bar{n} =
A P_{BE}(T) Q/Q_c$. Here, $A$ is the linear power loss from
additional cold attenuation, $P_{BE}=1/(e^{\hbar\omega/kT}-1)$ is
the Bose-Einstein population of the $50~\Omega$ load of the noise
source at effective temperature $T$, located outside the cavity.
The total cavity quality factor $Q =\omega \tau$ has an inverse
which is the sum of the inverses of the coupling quality factor
$Q_c$ of the noise source port, all other port couplings, and the
internal quality factor $Q_\text{int}$. In steady state and for
uncorrelated noise, the probability $P(N)$ of finding the qubit in
an environment with $N$ photons is a thermal distribution
$P(N)=\bar{n}^N / (\bar{n}+1)^{(N+1)}$, as verified by the data in
Fig.~\ref{rabiplot1}d. With these measurements we obtain the
scaling of $\bar{n}$ as a function of applied noise power for each
different value of $\kappa$, allowing a comparison with Eq.~2
using no adjustable parameters.


\begin{figure}
\includegraphics[width=3.375in]{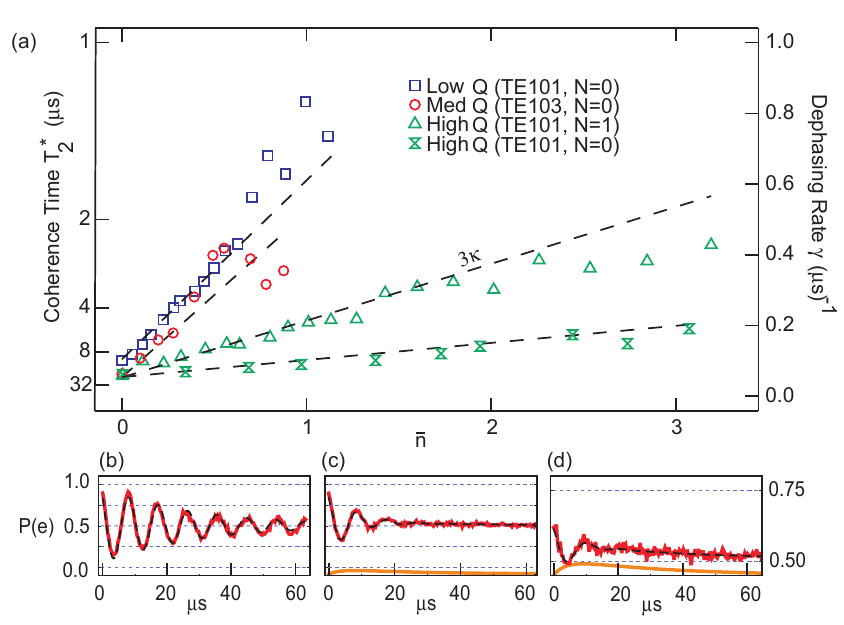} 
\caption{\label{fig1ver1} Qubit dephasing due to photon noise. (a)
Qubit coherence time, determined from Ramsey experiments on the
$N=0$ or $N=1$ ($\triangle$) photon peaks, as a function of both
cavity $Q$ and $\bar{n}$.  The dashed lines are theory, with an
offset due to residual dephasing.  Each has a slope proportional
to $\kappa$ (or $3\kappa$ for $N=1$ experiments), according to
Eq.~\ref{gammaout}. The ($\circ$) are coherence times vs.
population in $\text{TE}_{103}$ mode, which also dephases the
qubit. (b) Ramsey with no noise injected, fundamental mode
$Q=1\times10^6$, and $T_2^* = 26~\mu$s. The solid line is a fit
with an exponentially decaying sine. (c) A Ramsey with moderate
noise. Contrast and $T_2^*$ are reduced. Fundamental mode
$Q=2.5\times10^5$, $\bar{n}=0.25$, $T_2^*=7.7~\mu$s. (d) Ramsey
with high noise. Fundamental mode $Q=1\times10^6$, $\bar{n}=3.1$,
$T_2^*=5.2~\mu$s. Our selective ($N=0$) pulses produce a loss of
contrast and a non-oscillating signal addition (orange) as photon
population returns to a thermal distribution. The dashed black
line is a numerical simulation (see the Supplemental
Materials).}\label{gammaplot}
\end{figure}

To observe the influence of photon dephasing on our qubit, we test
Eq.~\ref{gammaout} over a wide range of values for  both $\bar{n}$
and $\kappa$ as shown in Fig.~\ref{gammaplot}.  The photon number
is varied by adjusting the attenuation following our noise source,
while $\kappa$ is controlled by retracting the resonator output
coupler using a Kevlar string connected to the top of the fridge,
exponentially increasing the $Q_c$ as it is withdrawn. For large
$\kappa$, photons enter and leave quickly, so long periods
uninterrupted by a transit are rare even if the average occupation
is low, and the phase coherence time is short.  In the Ramsey data
shown in Fig.~\ref{gammaplot} the dephasing rate is universally
proportional to injected $\bar{n}$ and $\kappa$, with an offset
due to spontaneous decay (if $N>0$), and residual photons or other
intrinsic dephasing. These experiments confirm our understanding
of the qubit dephasing rate in the strong-dispersive limit, and
point to the importance of excess photons or an effective
temperature of a mode for qubit coherence.

Importantly, we use slow Gaussian pulses to control the qubit in
order to exploit the photon-dependence of our Hamiltonian. With a
width of $\sigma=100$~ns, the narrow frequency span of the pulses
means that Ramsey experiments add signal contrast only when the
chosen photon number $N$ has remained in the cavity throughout the
experiment, a type of post-selection evident in the different
scalings of Fig.~\ref{gammaplot}b-d. Once conditioned, photon
transitions during the experiment lead to an incoherent response
in our qubit readout, when at a random point in time $t_0$ an
initially prepared superposition changes: $\ket{\psi(t_0)}
=1/\sqrt{2} \left(\ket{\text{g,0}} +
\ket{\text{e,0}}\right)\rightarrow \ket{\psi(t)}= 1/\sqrt{2}
\left(\ket{\text{g,1}} + \text{exp}[i \chi(t-t_0)]
\ket{\text{e,1}} \right)$ for time $t > t_0$.  Our qubit
readout~\cite{reed_high-fidelity_2010} traces over all photon
states, and the unknown final phase of the superposition produces
a decay in the Ramsey fringes, as the experiment records the qubit
excitation despite any cavity transition. Additionally, a
characteristic bump and slope are visible in the data and must be
removed before fitting the Ramsey signal with the usual decaying
sine function. These features can be understood as the
re-equilibration of the cavity photon number after the first qubit
manipulation conditionally prepares a certain photon number, and
are well fit (see Fig.~3 of the Supplement) by a simple master
equation which includes the incoherent cavity drive as well as
qubit and cavity decay.

While the fundamental $\text{TE}_{101}$ mode of our 3D resonator
serves both as the qubit readout channel and as a mechanism for
dephasing, the rectangular cavity in fact supports a set of
$\text{TE}_{10n}$ modes~\cite{poole_electron_1967} whose influence
we must consider. Then a more comprehensive Hamiltonian than Eq.~1
must incorporate many different cavity frequencies, each with a
coupling strength that depends on antenna length and the
positioning of the qubit in the cavity~\cite{nigg_BBQ_2012}.  This
coupling $g_n$ is large for odd-$n$ $\text{TE}_{10n}$ modes where
the electric field has an antinode at the qubit, while the
even-$n$ modes have greatly diminished coupling to the qubit due
to a node along the qubit antenna.  For our parameters, the
fundamental $\text{TE}_{101}$ mode $\omega_1/2\pi = 8.01$~GHz,
$\omega_q/2\pi = 6.65$~GHz, and $g_1/2\pi = 127$~MHz, the qubit
anharmonicity $\alpha=340$~MHz leads to an ac-Stark shift of
$\chi_1/2\pi = 7$~MHz.  Similarly, the first odd harmonics
$\text{TE}_{103}$ with $\omega_3 = 12.8$~GHz has a large
$\chi_3/2\pi = 1$~MHz.  In fact, with this mode we can perform
high fidelity readout, measure the photon mode population (using
longer $\sigma = 800$~ns width pulses), and observe its influence
on decoherence by injecting noise near $\omega_3$. In general, we
should consider {\it all} cavity modes that have a non-zero
coupling to the qubit as sources of significant decoherence. For
example, the odd-$n$ $\text{TE}_{10n}$ modes at frequency
$\omega_n$ and detuning $\Delta_n= \omega_n-\omega_q$, have a
coupling $g_{n} \propto \sqrt{\omega_n}$ and an ac-Stark shift
$\chi_n=g_n^2 \alpha/\Delta_n(\Delta_n+\alpha)$ which decrease
only slowly as $1/n$. Consequently, there may be many modes with
significant dispersive shifts that can act as sources of
extrinisic qubit decoherence. Moreover, since the coupling quality
factors of these modes typically decreases with frequency, even
very small photon occupancies (which are usually ignored, not
measured or as carefully filtered) must be suppressed to obtain
maximum coherence.

\begin{figure}
\includegraphics[width=3.375in]{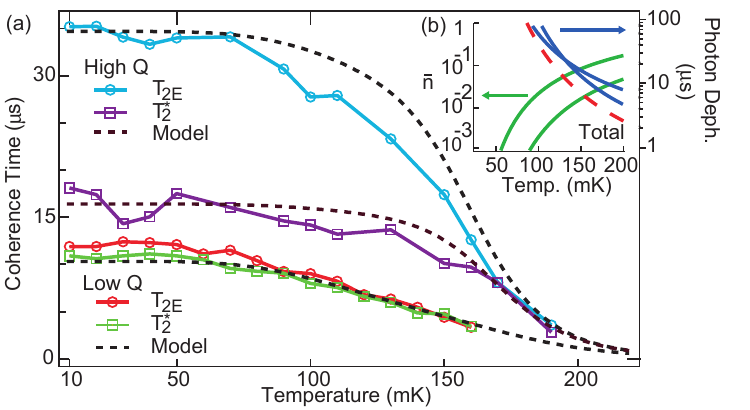} 
\caption{\label{tsweep} (a) Decoherence due to thermal photons.
The coherence times extracted from Ramsey ($T_2^*$) and Hahn echo
($T_\text{2E}$) experiments measured as a function of cryostat
temperature. To model dephasing (dashed lines), we predict
population in the $\text{TE}_{101}$ and $\text{TE}_{103}$ modes of
the cavity. Then, we sum the total dephasing rate using the
measured quality factors for each mode (High $Q$:
$\tau_{101}=20~\mu$s, $\tau_{103}=4~\mu$s; Low $Q$:
$\tau_{101}=2~\mu$s, $\tau_{103}=400$~ns). For high $Q$, the use
of a Hahn echo pulse leads to a large $T_{2E}$ because either the
photon state has much longer correlation time or the remaining
dephasing similarly occurs at low frequencies.  Although the
decline in $T_1$ (not shown)~\cite{catelani_quasiparticle_2011}
contributes to the trend, population in both $\text{TE}_{101}$ and
$\text{TE}_{103}$ are needed for a good fit. (b) Bose-Einstein
population of the first two odd-$n$ $\text{TE}_{10n}$ modes at 8
and 12.8~GHz (green) and the coherence limits they impose
individually (blue) and collectively (dashed red) for the low $Q$
values measured above. }\label{tsweep}
\end{figure}

The photon shot noise from multiple cavity modes provides a simple
explanation for the anomalous qubit dephasing previously
observed~\cite{paik_observation_2011} as a function of cryostat
temperature. In this case, each cavity mode should be populated
with the Bose-Einstein probability $P_{BE}$ and these thermal
photons can make an unintended measurement of the qubit,
disrupting phase-sensitive experiments. The predicted occupancies
for the $\text{TE}_{101}$ and $\text{TE}_{103}$ modes are shown
(green lines) in the inset of Fig.~3, along with their predicted
dephasing (blue lines). Having confirmed the dephasing rates for
all modes individually we can now combine the effect of all modes
that strongly couple to the qubit: $\gamma_\phi = \sum{\bar{n}_i
\kappa_i}$. This total thermal decoherence rate is shown as the
red dashed line in the inset of Fig.~3, for typical parameters.
Since these modes have $\hbar\omega_n\gg k_B T$, the predicted
dephasing time is in excess of 100 microseconds below 80~mK due to
the exponentially suppressed number of blackbody photons. However,
since any particular mode coupling to the qubit in the
strong-dispersive limit may have a relatively fast decay time
$\tau$, even very small ($\sim 10^{-3}-10^{-2}$)~non-thermal
populations $\bar{n}$ could easily satisfy $\bar{n}\kappa \gg
1/2T_1$, limiting the coherence through pure dephasing alone to
$T_2^*\approx1/\gamma_\phi=\tau/\bar{n}$. The measured coherence
times as a function of temperature are well fit (see Fig.~3) by
the combined dephasing of thermal occupancy of the
$\text{TE}_{101}$ and $\text{TE}_{103}$ modes, plus a parameter
adjusted to represent the residual dephasing in each experiment.
This excess could be due to another mechanism intrinsic to the
qubit, or simply due to insufficient filtering or thermalization
of the apparatus, leading to a small non-thermal photon
population.

\begin{figure}
\includegraphics[width=3.375in]{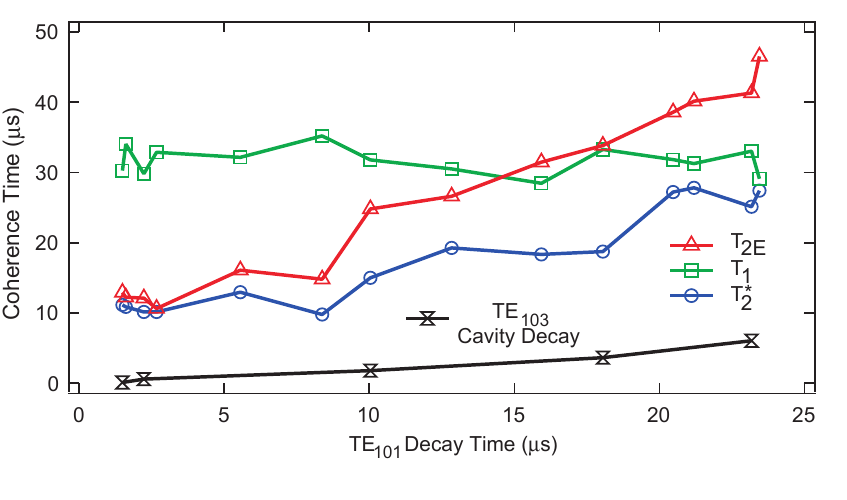} 
\caption{\label{pinplot} Coherence times versus  $\text{TE}_{101}$
mode decay $\tau$.  The $\text{TE}_{103}$ cavity, which naturally
decays more strongly through the couplers, increases in $Q$ as the
entire resonator is decoupled from our coaxial lines.  While $T_1$
is nearly constant due to the large qubit detuning from the
cavity, its $T_2^*$ and $T_{2E}$ increase as the coupling pin is
withdrawn from the 3D resonator. This is consistent with
diminishing dephasing from cavity modes with  $\kappa < \chi$,
where a photon transit strongly
measures~\cite{serban_crossover_2007} the qubit state.}
\label{pinplot}
\end{figure}


Further evidence that the true intrinsic coherence limits of the
3D transmons have not yet been observed is provided by the data
shown in Fig.~4, where the qubit relaxation time ($T_1$), Ramsey
time ($T_2^*$), and Hahn echo time ($T_\text{2E}$) at 10~mK are
shown as a function of the $\text{TE}_{101}$ cavity decay time. We
see that the relaxation time is relatively unaffected by cavity
lifetime, since this qubit is sufficiently detuned from the cavity
to minimize the Purcell effect~\cite{houck_controlling_2008}.
However, we observe a general trend where $T_2^*$ and $T_{2E}$
increase as the cavity lifetime increases. This is consistent with
a decoherence due to residual photons with ever slower dynamics,
but $\textit{not}$ expected due to e.g. junction critical current
noise, which should be independent of cavity properties.  The
maximum echo time (45~$\mu\rm{s}\approx 1.5~T_1$) observed here
indicates that coherence of small Josephson junction qubits is in
excess of several hundred microseconds when corrected by a single
Hahn echo.

In conclusion, we have performed experiments involving precise
thermal photon populations to quantitatively induce qubit
dephasing in good agreement with simple theory.  We find that
photons in the fundamental and at least one harmonic mode of the
cavity strongly couple to a transmon qubit and note that at the
nominal base temperature of our cryostat they produce a negligible
amount of dephasing. However, the sensitivity of the qubit to
photons at many frequencies requires that we either keep all modes
of the cavity in their ground state, or else minimize the
influence of non-thermal populations by reducing their measurement
rate~\cite{hatridgeshankar}. Inclusion of the cavity harmonics in
dephasing calculations leads to an understanding of the earlier,
anomalous, temperature-dependent decoherence in our
devices~\cite{paik_observation_2011}. Finally, we find evidence
that interactions with the residual photons in our 3D cavity
likely mask the intrinsic coherence time of the Josephson
junction, whose limits are much longer than qubit coherence times
seen so far.  As qubit linewidths shrink in the future, other
effects such as quasiparticle
parity~\cite{schreier_suppressing_2008, sun_measurements_2012,
naaman_time-domain_2006} or nuclear
spins~\cite{schuster_high-cooperativity_2010} may further split
the qubit spectrum, enabling probes of their state dynamics using
these procedures.

We thank Michel Devoret for valuable discussions.  L. F.
acknowledges partial support from CNR-Istituto di Cibernetica.
This research was funded by the Office of the Director of National
Intelligence (ODNI), Intelligence Advanced Research Projects
Activity (IARPA), through the Army Research Office, as well as by
the National Science Foundation (NSF DMR-1004406). All statements
of fact, opinion or conclusions contained herein are those of the
authors and should not be construed as representing the official
views or policies of IARPA, or the U.S. Government.

\begin{figure}[ht]
\centering
\includegraphics[width=7in]{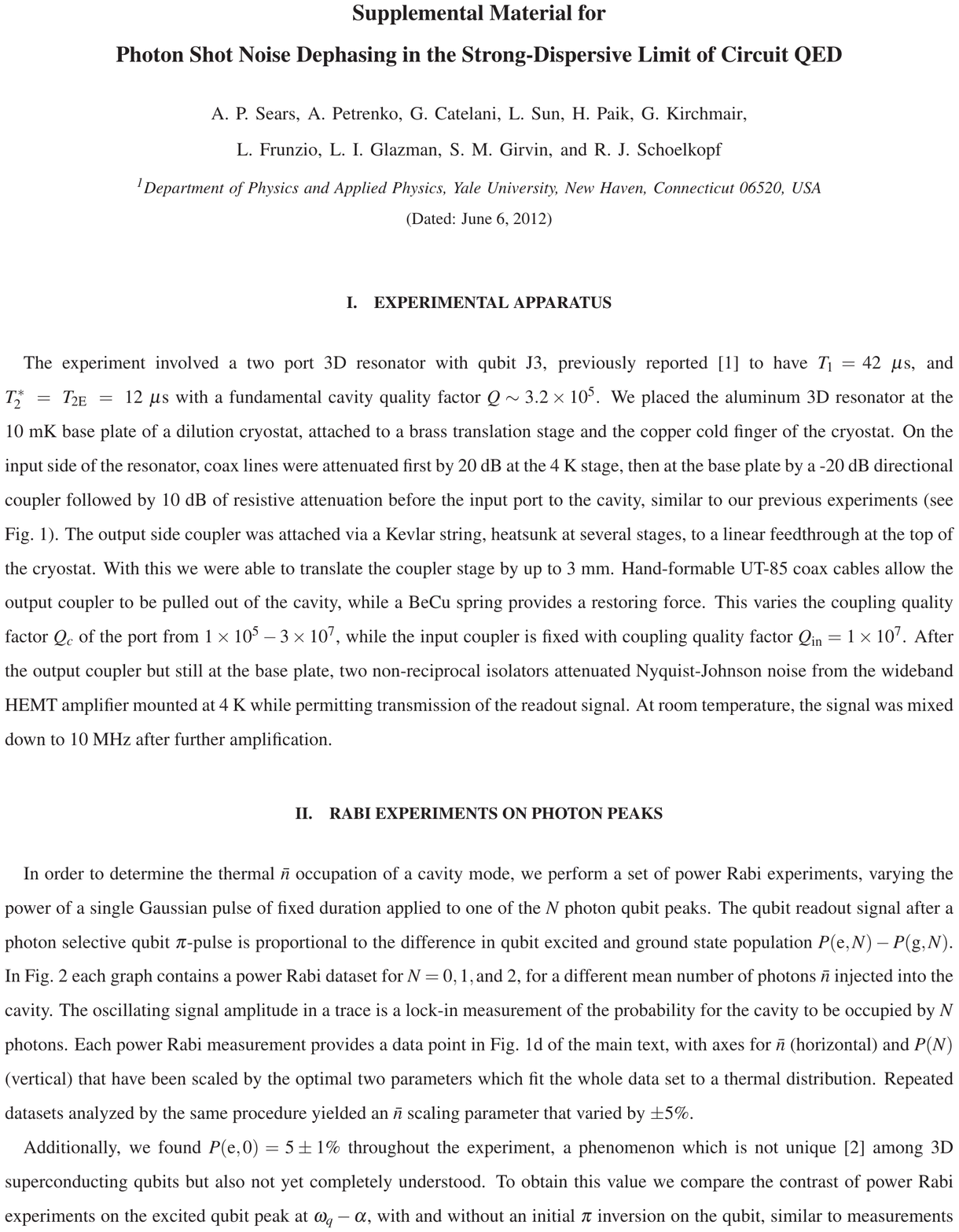}

\end{figure}

\begin{figure}[ht]
\centering
\includegraphics[width=7in]{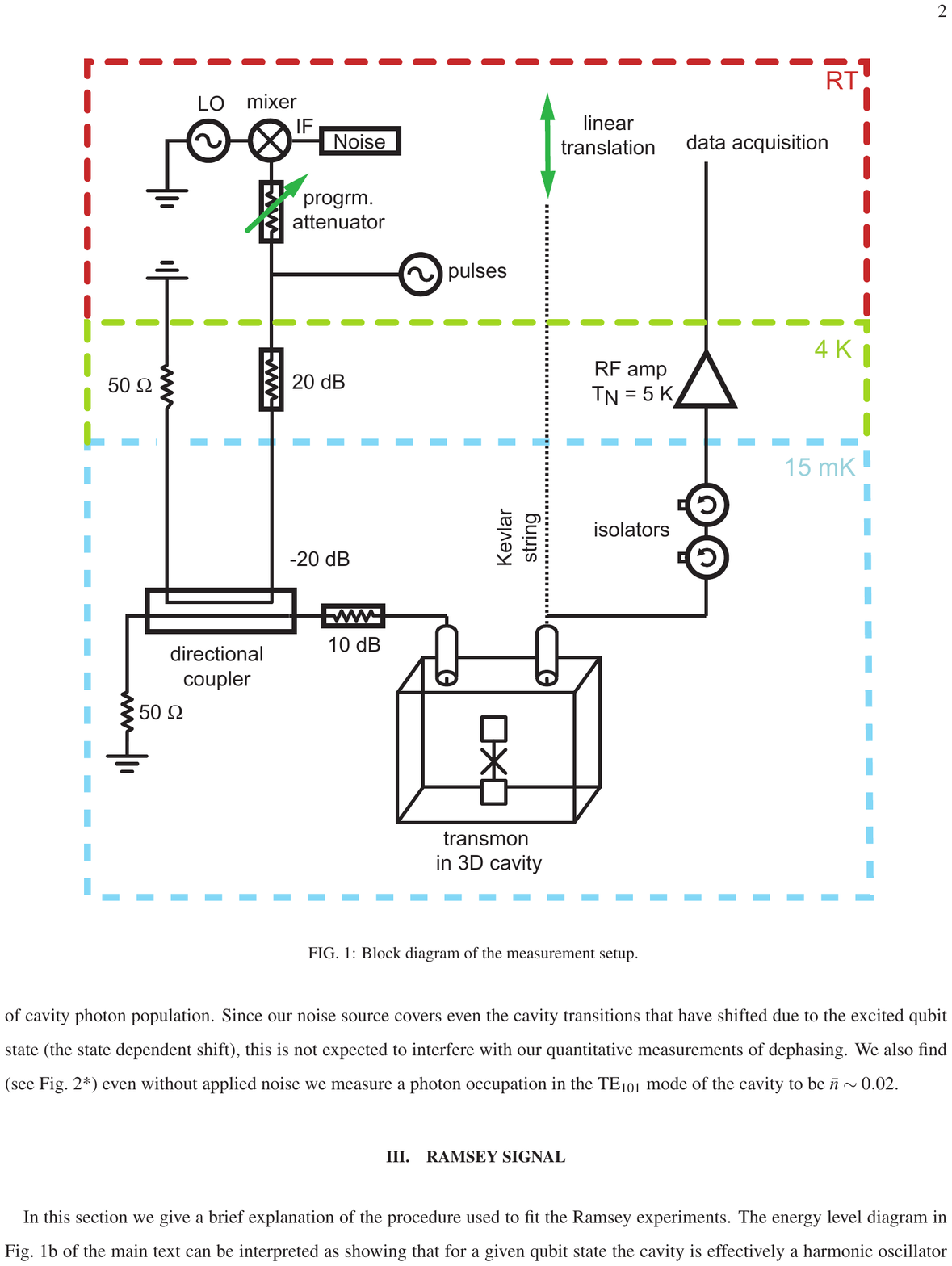}

\end{figure}

\begin{figure}[ht]
\centering
\includegraphics[width=7in]{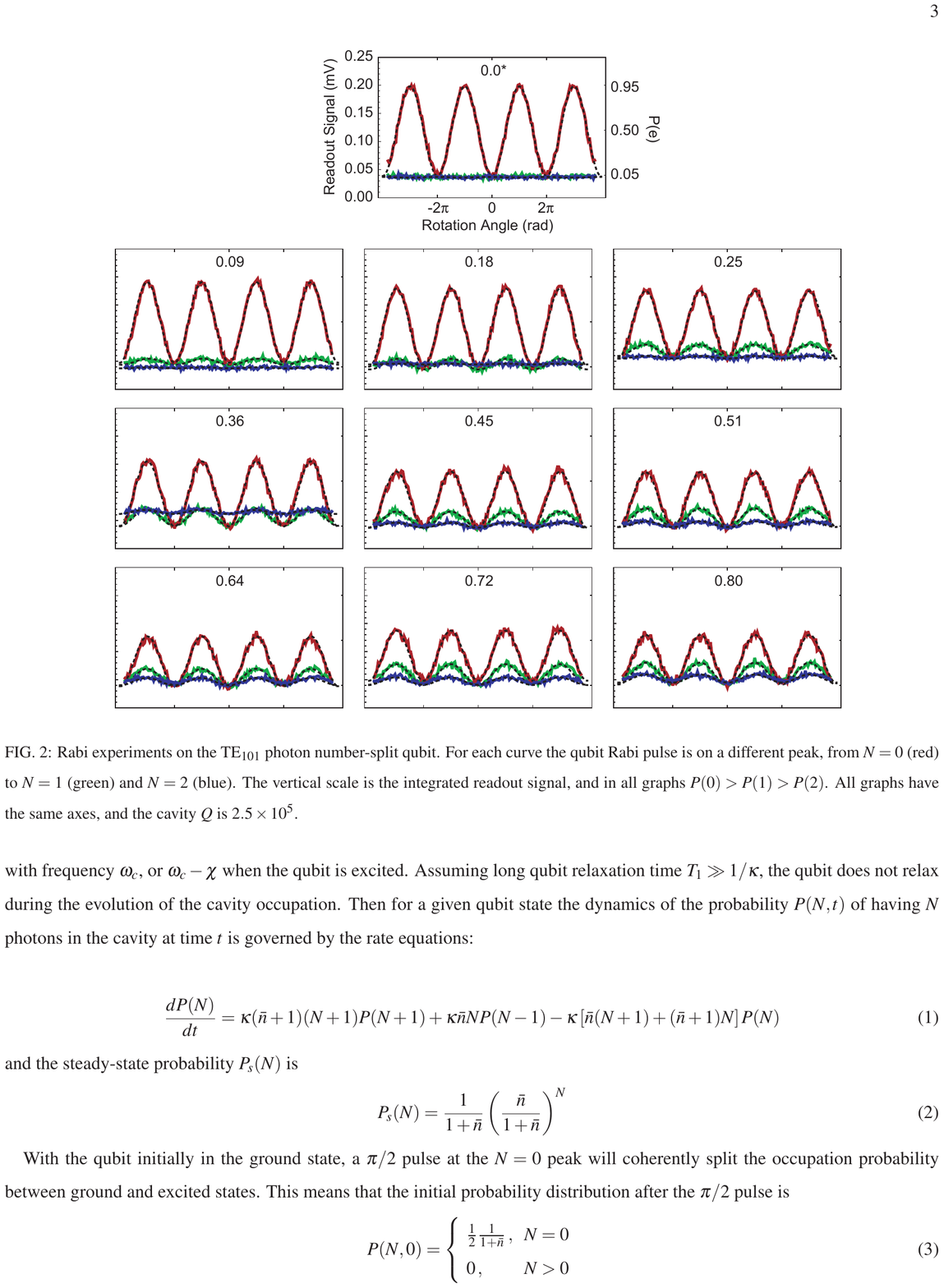}

\end{figure}

\begin{figure}[ht]
\centering
\includegraphics[width=7in]{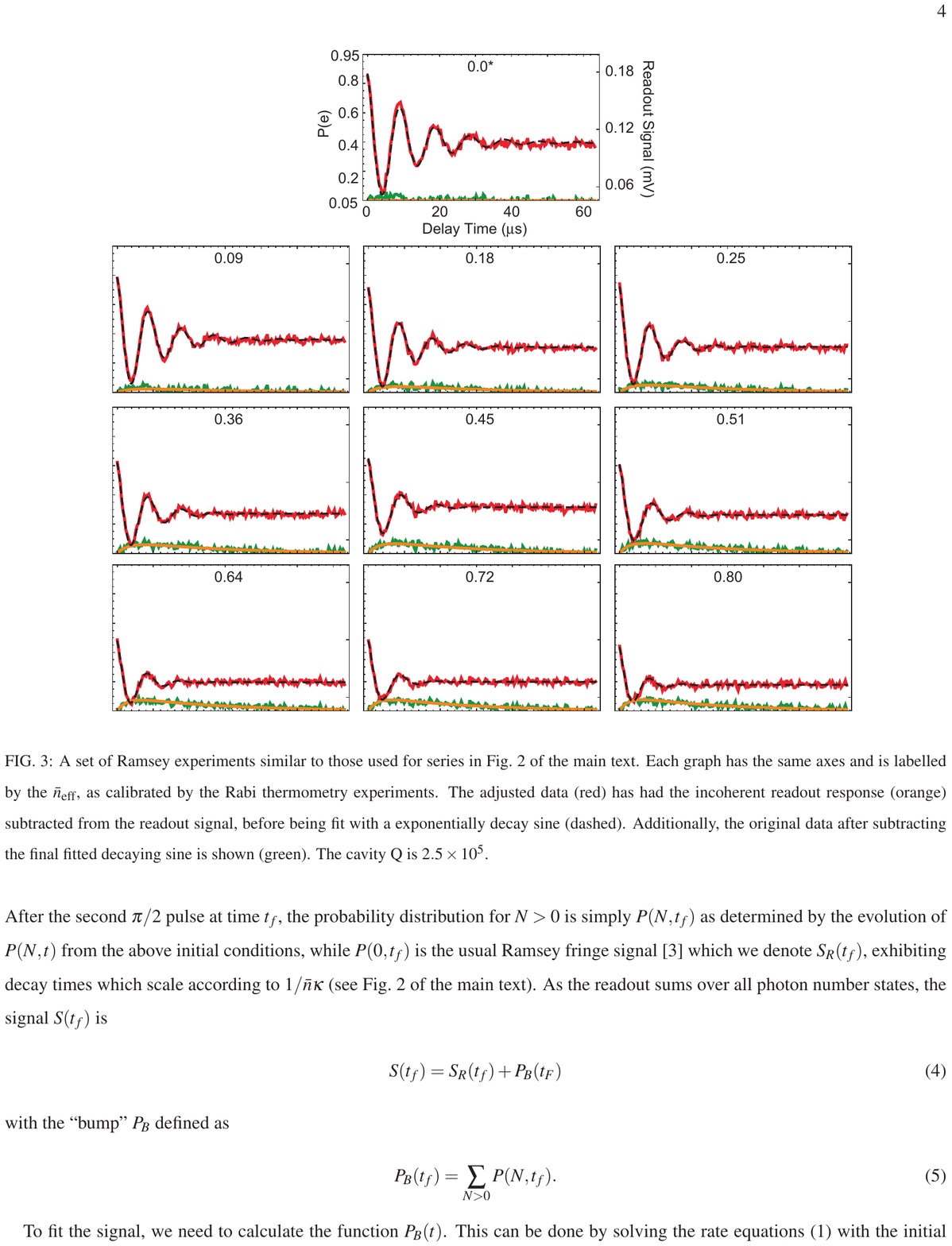}

\end{figure}

\begin{figure}[ht]
\centering
\includegraphics[width=7in]{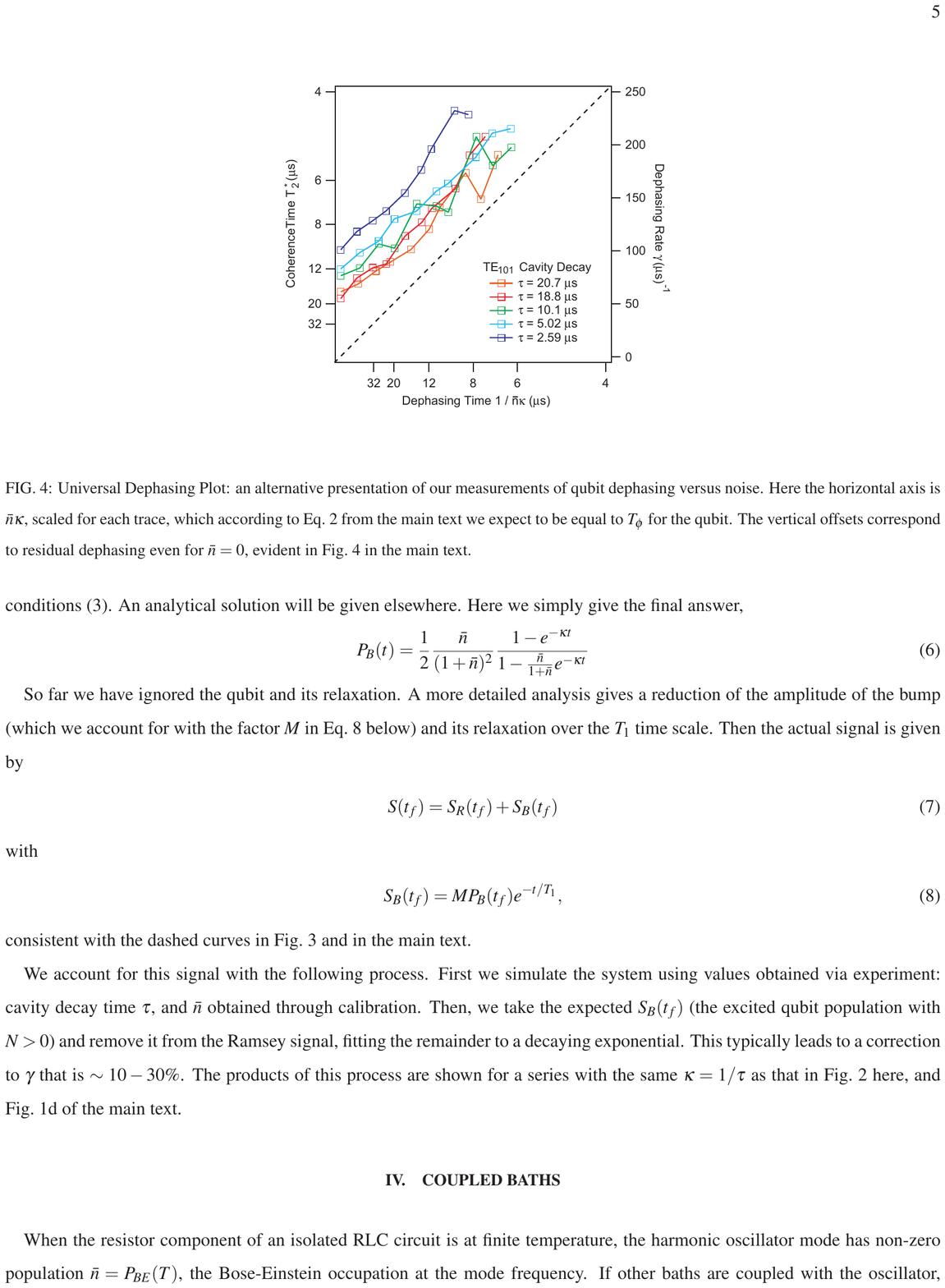}
\end{figure}

\begin{figure}[ht]
\centering
\includegraphics[width=7in]{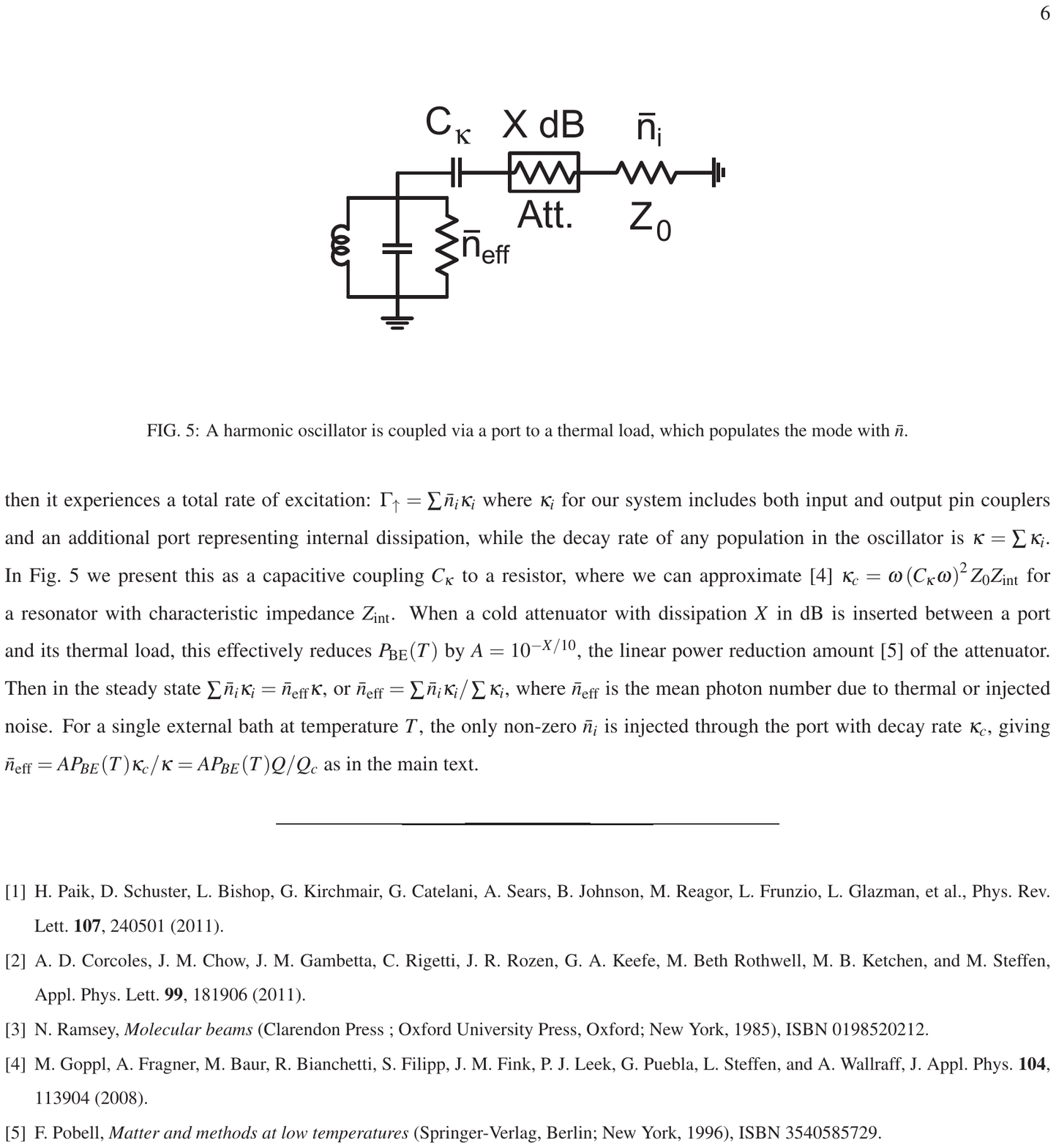}
\end{figure}


\begin{thebibliography}{24}
\expandafter\ifx\csname
natexlab\endcsname\relax\def\natexlab#1{#1}\fi
\expandafter\ifx\csname bibnamefont\endcsname\relax
  \def\bibnamefont#1{#1}\fi
\expandafter\ifx\csname bibfnamefont\endcsname\relax
  \def\bibfnamefont#1{#1}\fi
\expandafter\ifx\csname citenamefont\endcsname\relax
  \def\citenamefont#1{#1}\fi
\expandafter\ifx\csname url\endcsname\relax
  \def\url#1{\texttt{#1}}\fi
\expandafter\ifx\csname
urlprefix\endcsname\relax\def\urlprefix{URL }\fi
\providecommand{\bibinfo}[2]{#2}
\providecommand{\eprint}[2][]{\url{#2}}

\bibitem[{\citenamefont{Houck et~al.}(2008)\citenamefont{Houck, Schreier,
  Johnson, Chow, Koch, Gambetta, Schuster, Frunzio, Devoret, Girvin
  et~al.}}]{houck_controlling_2008}
\bibinfo{author}{\bibfnamefont{A.}~\bibnamefont{Houck}},
  \bibinfo{author}{\bibfnamefont{J.}~\bibnamefont{Schreier}},
  \bibinfo{author}{\bibfnamefont{B.}~\bibnamefont{Johnson}},
  \bibinfo{author}{\bibfnamefont{J.}~\bibnamefont{Chow}},
  \bibinfo{author}{\bibfnamefont{J.}~\bibnamefont{Koch}},
  \bibinfo{author}{\bibfnamefont{J.}~\bibnamefont{Gambetta}},
  \bibinfo{author}{\bibfnamefont{D.}~\bibnamefont{Schuster}},
  \bibinfo{author}{\bibfnamefont{L.}~\bibnamefont{Frunzio}},
  \bibinfo{author}{\bibfnamefont{M.}~\bibnamefont{Devoret}},
  \bibinfo{author}{\bibfnamefont{S.}~\bibnamefont{Girvin}},
  \bibnamefont{et~al.}, \bibinfo{journal}{Phys. Rev. Lett.}
  \textbf{\bibinfo{volume}{101}}, \bibinfo{pages}{080502}
  (\bibinfo{year}{2008}).

\bibitem[{\citenamefont{Neeley et~al.}(2008)\citenamefont{Neeley, Ansmann,
  Bialczak, Hofheinz, Katz, Lucero, {O'Connell}, Wang, Cleland, and
  Martinis}}]{neeley_transformed_2008}
\bibinfo{author}{\bibfnamefont{M.}~\bibnamefont{Neeley}},
  \bibinfo{author}{\bibfnamefont{M.}~\bibnamefont{Ansmann}},
  \bibinfo{author}{\bibfnamefont{R.}~\bibnamefont{Bialczak}},
  \bibinfo{author}{\bibfnamefont{M.}~\bibnamefont{Hofheinz}},
  \bibinfo{author}{\bibfnamefont{N.}~\bibnamefont{Katz}},
  \bibinfo{author}{\bibfnamefont{E.}~\bibnamefont{Lucero}},
  \bibinfo{author}{\bibfnamefont{A.}~\bibnamefont{{O'Connell}}},
  \bibinfo{author}{\bibfnamefont{H.}~\bibnamefont{Wang}},
  \bibinfo{author}{\bibfnamefont{A. N.}~\bibnamefont{Cleland}}, \bibnamefont{and}
  \bibinfo{author}{\bibfnamefont{J. M.}~\bibnamefont{Martinis}},
  \bibinfo{journal}{Phys. Rev. B} \textbf{\bibinfo{volume}{77}},
  \bibinfo{pages}{180508(R)} (\bibinfo{year}{2008}).

\bibitem[{\citenamefont{Steffen et~al.}(2009)\citenamefont{Steffen, Brito,
  {DiVincenzo}, Kumar, and Ketchen}}]{steffen_decoherence_2009}
\bibinfo{author}{\bibfnamefont{M.}~\bibnamefont{Steffen}},
  \bibinfo{author}{\bibfnamefont{F.}~\bibnamefont{Brito}},
  \bibinfo{author}{\bibfnamefont{D.}~\bibnamefont{{DiVincenzo}}},
  \bibinfo{author}{\bibfnamefont{S.}~\bibnamefont{Kumar}}, \bibnamefont{and}
  \bibinfo{author}{\bibfnamefont{M.}~\bibnamefont{Ketchen}},
  \bibinfo{journal}{New J. Phys.} \textbf{\bibinfo{volume}{11}},
  \bibinfo{pages}{033030} (\bibinfo{year}{2009}).

\bibitem[{\citenamefont{Paik et~al.}(2011)\citenamefont{Paik, Schuster, Bishop,
  Kirchmair, Catelani, Sears, Johnson, Reagor, Frunzio, Glazman
  et~al.}}]{paik_observation_2011}
\bibinfo{author}{\bibfnamefont{H.}~\bibnamefont{Paik}},
  \bibinfo{author}{\bibfnamefont{D.}~\bibnamefont{Schuster}},
  \bibinfo{author}{\bibfnamefont{L.}~\bibnamefont{Bishop}},
  \bibinfo{author}{\bibfnamefont{G.}~\bibnamefont{Kirchmair}},
  \bibinfo{author}{\bibfnamefont{G.}~\bibnamefont{Catelani}},
  \bibinfo{author}{\bibfnamefont{A.}~\bibnamefont{Sears}},
  \bibinfo{author}{\bibfnamefont{B.}~\bibnamefont{Johnson}},
  \bibinfo{author}{\bibfnamefont{M.}~\bibnamefont{Reagor}},
  \bibinfo{author}{\bibfnamefont{L.}~\bibnamefont{Frunzio}},
  \bibinfo{author}{\bibfnamefont{L.}~\bibnamefont{Glazman}},
  \bibnamefont{et~al.}, \bibinfo{journal}{Phys. Rev. Lett.}
  \textbf{\bibinfo{volume}{107}}, \bibinfo{pages}{240501}
  (\bibinfo{year}{2011}).

\bibitem[{\citenamefont{Houck et~al.}(2009)\citenamefont{Houck, Koch, Devoret,
  Girvin, and Schoelkopf}}]{houck_life_2009}
\bibinfo{author}{\bibfnamefont{A.~A.} \bibnamefont{Houck}},
  \bibinfo{author}{\bibfnamefont{J.}~\bibnamefont{Koch}},
  \bibinfo{author}{\bibfnamefont{M.~H.} \bibnamefont{Devoret}},
  \bibinfo{author}{\bibfnamefont{S.~M.} \bibnamefont{Girvin}},
  \bibnamefont{and} \bibinfo{author}{\bibfnamefont{R.~J.}
  \bibnamefont{Schoelkopf}}, \bibinfo{journal}{Quantum Inf. Process.}
  \textbf{\bibinfo{volume}{8}}, \bibinfo{pages}{105} (\bibinfo{year}{2009}).

\bibitem[{\citenamefont{Wellstood et~al.}(1987)\citenamefont{Wellstood, Urbina,
  and Clarke}}]{wellstood_low-frequency_1987}
\bibinfo{author}{\bibfnamefont{F.~C.} \bibnamefont{Wellstood}},
  \bibinfo{author}{\bibfnamefont{C.}~\bibnamefont{Urbina}}, \bibnamefont{and}
  \bibinfo{author}{\bibfnamefont{J.}~\bibnamefont{Clarke}},
  \bibinfo{journal}{Appl. Phys. Lett.} \textbf{\bibinfo{volume}{50}},
  \bibinfo{pages}{772} (\bibinfo{year}{1987}).

\bibitem[{\citenamefont{Van~Harlingen et~al.}(2004)\citenamefont{Van~Harlingen,
  Robertson, Plourde, Reichardt, Crane, and
  Clarke}}]{van_harlingen_decoherence_2004}
\bibinfo{author}{\bibfnamefont{D.}~\bibnamefont{Van~Harlingen}},
  \bibinfo{author}{\bibfnamefont{T.}~\bibnamefont{Robertson}},
  \bibinfo{author}{\bibfnamefont{B.}~\bibnamefont{Plourde}},
  \bibinfo{author}{\bibfnamefont{P.}~\bibnamefont{Reichardt}},
  \bibinfo{author}{\bibfnamefont{T.}~\bibnamefont{Crane}}, \bibnamefont{and}
  \bibinfo{author}{\bibfnamefont{J.}~\bibnamefont{Clarke}},
  \bibinfo{journal}{Phys. Rev. B} \textbf{\bibinfo{volume}{70}},
  \bibinfo{pages}{064517} (\bibinfo{year}{2004}).

\bibitem[{\citenamefont{Eroms et~al.}(2006)\citenamefont{Eroms, van
  Schaarenburg, Driessen, Plantenberg, Huizinga, Schouten, Verbruggen, Harmans,
  and Mooij}}]{eroms_low-frequency_2006}
\bibinfo{author}{\bibfnamefont{J.}~\bibnamefont{Eroms}},
  \bibinfo{author}{\bibfnamefont{L.~C.} \bibnamefont{van Schaarenburg}},
  \bibinfo{author}{\bibfnamefont{E.~F.~C.} \bibnamefont{Driessen}},
  \bibinfo{author}{\bibfnamefont{J.~H.} \bibnamefont{Plantenberg}},
  \bibinfo{author}{\bibfnamefont{C.~M.} \bibnamefont{Huizinga}},
  \bibinfo{author}{\bibfnamefont{R.~N.} \bibnamefont{Schouten}},
  \bibinfo{author}{\bibfnamefont{A.~H.} \bibnamefont{Verbruggen}},
  \bibinfo{author}{\bibfnamefont{C.~J. P.~M.} \bibnamefont{Harmans}},
  \bibnamefont{and} \bibinfo{author}{\bibfnamefont{J.~E.} \bibnamefont{Mooij}},
  \bibinfo{journal}{Appl. Phys. Lett.} \textbf{\bibinfo{volume}{89}},
  \bibinfo{pages}{122516} (\bibinfo{year}{2006}).

\bibitem[{\citenamefont{Bertet et~al.}(2005)\citenamefont{Bertet, Chiorescu,
  Burkard, Semba, Harmans, {DiVincenzo}, and Mooij}}]{bertet_dephasing_2005}
\bibinfo{author}{\bibfnamefont{P.}~\bibnamefont{Bertet}},
  \bibinfo{author}{\bibfnamefont{I.}~\bibnamefont{Chiorescu}},
  \bibinfo{author}{\bibfnamefont{G.}~\bibnamefont{Burkard}},
  \bibinfo{author}{\bibfnamefont{K.}~\bibnamefont{Semba}},
  \bibinfo{author}{\bibfnamefont{C.}~\bibnamefont{Harmans}},
  \bibinfo{author}{\bibfnamefont{D.}~\bibnamefont{{DiVincenzo}}},
  \bibnamefont{and} \bibinfo{author}{\bibfnamefont{J.}~\bibnamefont{Mooij}},
  \bibinfo{journal}{Phys. Rev. Lett.} \textbf{\bibinfo{volume}{95}},
  \bibinfo{pages}{257002} (\bibinfo{year}{2005}).

\bibitem[{\citenamefont{Gambetta et~al.}(2006)\citenamefont{Gambetta, Blais,
  Schuster, Wallraff, Frunzio, Majer, Devoret, Girvin, and
  Schoelkopf}}]{gambetta_qubit-photon_2006}
\bibinfo{author}{\bibfnamefont{J.}~\bibnamefont{Gambetta}},
  \bibinfo{author}{\bibfnamefont{A.}~\bibnamefont{Blais}},
  \bibinfo{author}{\bibfnamefont{D.}~\bibnamefont{Schuster}},
  \bibinfo{author}{\bibfnamefont{A.}~\bibnamefont{Wallraff}},
  \bibinfo{author}{\bibfnamefont{L.}~\bibnamefont{Frunzio}},
  \bibinfo{author}{\bibfnamefont{J.}~\bibnamefont{Majer}},
  \bibinfo{author}{\bibfnamefont{M.}~\bibnamefont{Devoret}},
  \bibinfo{author}{\bibfnamefont{S.}~\bibnamefont{Girvin}}, \bibnamefont{and}
  \bibinfo{author}{\bibfnamefont{R.}~\bibnamefont{Schoelkopf}},
  \bibinfo{journal}{Phys. Rev. A} \textbf{\bibinfo{volume}{74}},
  \bibinfo{pages}{042318} (\bibinfo{year}{2006}).

\bibitem[{\citenamefont{Serban et~al.}(2007)\citenamefont{Serban, Solano, and
  Wilhelm}}]{serban_crossover_2007}
\bibinfo{author}{\bibfnamefont{I.}~\bibnamefont{Serban}},
  \bibinfo{author}{\bibfnamefont{E.}~\bibnamefont{Solano}}, \bibnamefont{and}
  \bibinfo{author}{\bibfnamefont{F.~K.} \bibnamefont{Wilhelm}},
  \bibinfo{journal}{Europhysics Letters {(EPL)}} \textbf{\bibinfo{volume}{80}},
  \bibinfo{pages}{40011} (\bibinfo{year}{2007}).

\bibitem[{\citenamefont{Schuster et~al.}(2005)\citenamefont{Schuster, Wallraff,
  Blais, Frunzio, Huang, Majer, Girvin, and Schoelkopf}}]{schuster_ac_2005}
\bibinfo{author}{\bibfnamefont{D.}~\bibnamefont{Schuster}},
  \bibinfo{author}{\bibfnamefont{A.}~\bibnamefont{Wallraff}},
  \bibinfo{author}{\bibfnamefont{A.}~\bibnamefont{Blais}},
  \bibinfo{author}{\bibfnamefont{L.}~\bibnamefont{Frunzio}},
  \bibinfo{author}{\bibfnamefont{R.}~\bibnamefont{Huang}},
  \bibinfo{author}{\bibfnamefont{J.}~\bibnamefont{Majer}},
  \bibinfo{author}{\bibfnamefont{S.}~\bibnamefont{Girvin}}, \bibnamefont{and}
  \bibinfo{author}{\bibfnamefont{R.}~\bibnamefont{Schoelkopf}},
  \bibinfo{journal}{Phys. Rev. Lett.} \textbf{\bibinfo{volume}{94}},
  \bibinfo{pages}{123602} (\bibinfo{year}{2005}).

\bibitem[{\citenamefont{Rigetti et~al.}(2012)\citenamefont{Rigetti, Poletto,
  Gambetta, Plourde, Chow, Corcoles, Smolin, Merkel, Rozen, Keefe
  et~al.}}]{rigetti_superconducting_2012}
\bibinfo{author}{\bibfnamefont{C.}~\bibnamefont{Rigetti}},
  \bibinfo{author}{\bibfnamefont{S.}~\bibnamefont{Poletto}},
  \bibinfo{author}{\bibfnamefont{J.}~\bibnamefont{Gambetta}},
  \bibinfo{author}{\bibfnamefont{B.}~\bibnamefont{Plourde}},
  \bibinfo{author}{\bibfnamefont{J.}~\bibnamefont{Chow}},
  \bibinfo{author}{\bibfnamefont{A.}~\bibnamefont{Corcoles}},
  \bibinfo{author}{\bibfnamefont{J.}~\bibnamefont{Smolin}},
  \bibinfo{author}{\bibfnamefont{S.}~\bibnamefont{Merkel}},
  \bibinfo{author}{\bibfnamefont{J.}~\bibnamefont{Rozen}},
  \bibinfo{author}{\bibfnamefont{G.}~\bibnamefont{Keefe}},
  \bibnamefont{et~al.}, \bibinfo{journal}{Arxiv preprint {arXiv:1202.5533}}
  (\bibinfo{year}{2012}).

\bibitem[{\citenamefont{Sun et~al.}(2010)\citenamefont{L. Sun and A. Petrenko}}]{luyanandrei}
\bibinfo{author}{\bibfnamefont{L.}~\bibnamefont{Sun}} \bibnamefont{and}
\bibinfo{author}{\bibfnamefont{A.}~\bibnamefont{Petrenko}},
\bibinfo{note}{in prep}.

\bibitem[{\citenamefont{Schuster et~al.}(2007)\citenamefont{Schuster, Houck,
  Schreier, Wallraff, Gambetta, Blais, Frunzio, Majer, Johnson, Devoret
  et~al.}}]{Schuster_resolving_2007}
\bibinfo{author}{\bibfnamefont{D.~I.} \bibnamefont{Schuster}},
  \bibinfo{author}{\bibfnamefont{A.~A.} \bibnamefont{Houck}},
  \bibinfo{author}{\bibfnamefont{J.~A.} \bibnamefont{Schreier}},
  \bibinfo{author}{\bibfnamefont{A.}~\bibnamefont{Wallraff}},
  \bibinfo{author}{\bibfnamefont{J.~M.} \bibnamefont{Gambetta}},
  \bibinfo{author}{\bibfnamefont{A.}~\bibnamefont{Blais}},
  \bibinfo{author}{\bibfnamefont{L.}~\bibnamefont{Frunzio}},
  \bibinfo{author}{\bibfnamefont{J.}~\bibnamefont{Majer}},
  \bibinfo{author}{\bibfnamefont{B.}~\bibnamefont{Johnson}},
  \bibinfo{author}{\bibfnamefont{M.~H.} \bibnamefont{Devoret}},
  \bibnamefont{et~al.}, \bibinfo{journal}{Nature}
  \textbf{\bibinfo{volume}{445}}, \bibinfo{pages}{515} (\bibinfo{year}{2007}).

\bibitem[{\citenamefont{Johnson et~al.}(2010)\citenamefont{Johnson, Reed,
  Houck, Schuster, Bishop, Ginossar, Gambetta, {DiCarlo}, Frunzio, Girvin
  et~al.}}]{johnson_quantum_2010}
\bibinfo{author}{\bibfnamefont{B.~R.} \bibnamefont{Johnson}},
  \bibinfo{author}{\bibfnamefont{M.~D.} \bibnamefont{Reed}},
  \bibinfo{author}{\bibfnamefont{A.~A.} \bibnamefont{Houck}},
  \bibinfo{author}{\bibfnamefont{D.~I.} \bibnamefont{Schuster}},
  \bibinfo{author}{\bibfnamefont{L.~S.} \bibnamefont{Bishop}},
  \bibinfo{author}{\bibfnamefont{E.}~\bibnamefont{Ginossar}},
  \bibinfo{author}{\bibfnamefont{J.~M.} \bibnamefont{Gambetta}},
  \bibinfo{author}{\bibfnamefont{L.}~\bibnamefont{{DiCarlo}}},
  \bibinfo{author}{\bibfnamefont{L.}~\bibnamefont{Frunzio}},
  \bibinfo{author}{\bibfnamefont{S.~M.} \bibnamefont{Girvin}},
  \bibnamefont{et~al.}, \bibinfo{journal}{Nature Phys.}
  \textbf{\bibinfo{volume}{6}}, \bibinfo{pages}{663} (\bibinfo{year}{2010}).

\bibitem[{\citenamefont{Reed et~al.}(2010)\citenamefont{Reed, {DiCarlo},
  Johnson, Sun, Schuster, Frunzio, and Schoelkopf}}]{reed_high-fidelity_2010}
\bibinfo{author}{\bibfnamefont{M.}~\bibnamefont{Reed}},
  \bibinfo{author}{\bibfnamefont{L.}~\bibnamefont{{DiCarlo}}},
  \bibinfo{author}{\bibfnamefont{B.}~\bibnamefont{Johnson}},
  \bibinfo{author}{\bibfnamefont{L.}~\bibnamefont{Sun}},
  \bibinfo{author}{\bibfnamefont{D.}~\bibnamefont{Schuster}},
  \bibinfo{author}{\bibfnamefont{L.}~\bibnamefont{Frunzio}}, \bibnamefont{and}
  \bibinfo{author}{\bibfnamefont{R.}~\bibnamefont{Schoelkopf}},
  \bibinfo{journal}{Phys. Rev. Lett.} \textbf{\bibinfo{volume}{105}},
  \bibinfo{pages}{173601} (\bibinfo{year}{2010}).

\bibitem[{\citenamefont{Nigg et~al.}(2012)\citenamefont{Nigg, Paik, Vlastakis,
  Kirchmair, Shankar, Frunzio, Devoret, Schoelkopf, and
  Girvin}}]{nigg_BBQ_2012}
\bibinfo{author}{\bibfnamefont{S.}~\bibnamefont{Nigg}},
  \bibinfo{author}{\bibfnamefont{H.}~\bibnamefont{Paik}},
  \bibinfo{author}{\bibfnamefont{B.}~\bibnamefont{Vlastakis}},
  \bibinfo{author}{\bibfnamefont{G.}~\bibnamefont{Kirchmair}},
  \bibinfo{author}{\bibfnamefont{S.}~\bibnamefont{Shankar}},
  \bibinfo{author}{\bibfnamefont{L.}~\bibnamefont{Frunzio}},
  \bibinfo{author}{\bibfnamefont{M.}~\bibnamefont{Devoret}},
  \bibinfo{author}{\bibfnamefont{R.}~\bibnamefont{Schoelkopf}},
  \bibnamefont{and} \bibinfo{author}{\bibfnamefont{S.}~\bibnamefont{Girvin}},
  \bibinfo{howpublished}{Arxiv preprint:1204.0587}
  (\bibinfo{year}{2012}), \bibinfo{note}{Phys. Rev. Lett. in press}.

\bibitem[{\citenamefont{Walls and Milburn}(1994)}]{walls_quantum_1994}
\bibinfo{author}{\bibfnamefont{D.~F.} \bibnamefont{Walls}} \bibnamefont{and}
  \bibinfo{author}{\bibfnamefont{G.~J.} \bibnamefont{Milburn}},
  \emph{\bibinfo{title}{Quantum optics}} (\bibinfo{publisher}{Springer},
  \bibinfo{address}{Berlin; New York}, \bibinfo{year}{1994}), ISBN
  \bibinfo{isbn}{3540588310}.

\bibitem[{\citenamefont{Poole}(1967)}]{poole_electron_1967}
\bibinfo{author}{\bibfnamefont{C.~P.} \bibnamefont{Poole}},
  \emph{\bibinfo{title}{Electron spin resonance : a comprehensive treatise on
  experimental techniques}} (\bibinfo{publisher}{Wiley}, \bibinfo{address}{New
  York [u.a.]}, \bibinfo{year}{1967}), ISBN \bibinfo{isbn}{9780470693865}.

\bibitem[{\citenamefont{Catelani et~al.}(2011)\citenamefont{Catelani, Koch,
  Frunzio, Schoelkopf, Devoret, and Glazman}}]{catelani_quasiparticle_2011}
\bibinfo{author}{\bibfnamefont{G.}~\bibnamefont{Catelani}},
  \bibinfo{author}{\bibfnamefont{J.}~\bibnamefont{Koch}},
  \bibinfo{author}{\bibfnamefont{L.}~\bibnamefont{Frunzio}},
  \bibinfo{author}{\bibfnamefont{R.}~\bibnamefont{Schoelkopf}},
  \bibinfo{author}{\bibfnamefont{M.}~\bibnamefont{Devoret}}, \bibnamefont{and}
  \bibinfo{author}{\bibfnamefont{L.}~\bibnamefont{Glazman}},
  \bibinfo{journal}{Phys. Rev. Lett.} \textbf{\bibinfo{volume}{106}},
  \bibinfo{pages}{077002} (\bibinfo{year}{2011}).

\bibitem[{\citenamefont{Hatridge et~al.}(2010)\citenamefont{M. Hatridge and S. Shankar}}]{hatridgeshankar}
\bibinfo{author}{\bibfnamefont{M.}~\bibnamefont{Hatridge}} \bibnamefont{and}
\bibinfo{author}{\bibfnamefont{S.}~\bibnamefont{Shankar}},
\bibinfo{note}{in prep}.


\bibitem[{\citenamefont{Schreier et~al.}(2008)\citenamefont{Schreier, Houck,
  Koch, Schuster, Johnson, Chow, Gambetta, Majer, Frunzio, Devoret
  et~al.}}]{schreier_suppressing_2008}
\bibinfo{author}{\bibfnamefont{J.}~\bibnamefont{Schreier}},
  \bibinfo{author}{\bibfnamefont{A.}~\bibnamefont{Houck}},
  \bibinfo{author}{\bibfnamefont{J.}~\bibnamefont{Koch}},
  \bibinfo{author}{\bibfnamefont{D.}~\bibnamefont{Schuster}},
  \bibinfo{author}{\bibfnamefont{B.}~\bibnamefont{Johnson}},
  \bibinfo{author}{\bibfnamefont{J.}~\bibnamefont{Chow}},
  \bibinfo{author}{\bibfnamefont{J.}~\bibnamefont{Gambetta}},
  \bibinfo{author}{\bibfnamefont{J.}~\bibnamefont{Majer}},
  \bibinfo{author}{\bibfnamefont{L.}~\bibnamefont{Frunzio}},
  \bibinfo{author}{\bibfnamefont{M.}~\bibnamefont{Devoret}},
  \bibnamefont{et~al.}, \bibinfo{journal}{Phys. Rev. B}
  \textbf{\bibinfo{volume}{77}}, \bibinfo{pages}{180502(R)}
  (\bibinfo{year}{2008}).


\bibitem[{\citenamefont{Sun et~al.}(2012)\citenamefont{Sun, {DiCarlo}, Reed,
  Catelani, Bishop, Schuster, Johnson, Yang, Frunzio, Glazman
  et~al.}}]{sun_measurements_2012}
\bibinfo{author}{\bibfnamefont{L.}~\bibnamefont{Sun}},
  \bibinfo{author}{\bibfnamefont{L.}~\bibnamefont{{DiCarlo}}},
  \bibinfo{author}{\bibfnamefont{M.}~\bibnamefont{Reed}},
  \bibinfo{author}{\bibfnamefont{G.}~\bibnamefont{Catelani}},
  \bibinfo{author}{\bibfnamefont{L.~S.} \bibnamefont{Bishop}},
  \bibinfo{author}{\bibfnamefont{D.~I.} \bibnamefont{Schuster}},
  \bibinfo{author}{\bibfnamefont{B.~R.} \bibnamefont{Johnson}},
  \bibinfo{author}{\bibfnamefont{G.~A.} \bibnamefont{Yang}},
  \bibinfo{author}{\bibfnamefont{L.}~\bibnamefont{Frunzio}},
  \bibinfo{author}{\bibfnamefont{L.}~\bibnamefont{Glazman}},
  \bibnamefont{et~al.},
  \bibinfo{howpublished}{Arxiv preprint:1112.2621}
  (\bibinfo{year}{2012}), \bibinfo{note}{Phys. Rev. Lett. in press}.


\bibitem[{\citenamefont{Naaman and Aumentado}(2006)}]{naaman_time-domain_2006}
\bibinfo{author}{\bibfnamefont{O.}~\bibnamefont{Naaman}} \bibnamefont{and}
  \bibinfo{author}{\bibfnamefont{J.}~\bibnamefont{Aumentado}},
  \bibinfo{journal}{Phys. Rev. B} \textbf{\bibinfo{volume}{73}},
  \bibinfo{pages}{172504} (\bibinfo{year}{2006}).

\bibitem[{\citenamefont{Schuster et~al.}(2010)\citenamefont{Schuster, Sears,
  Ginossar, {DiCarlo}, Frunzio, Morton, Wu, Briggs, Buckley, Awschalom
  et~al.}}]{schuster_high-cooperativity_2010}
\bibinfo{author}{\bibfnamefont{D.}~\bibnamefont{Schuster}},
  \bibinfo{author}{\bibfnamefont{A.}~\bibnamefont{Sears}},
  \bibinfo{author}{\bibfnamefont{E.}~\bibnamefont{Ginossar}},
  \bibinfo{author}{\bibfnamefont{L.}~\bibnamefont{{DiCarlo}}},
  \bibinfo{author}{\bibfnamefont{L.}~\bibnamefont{Frunzio}},
  \bibinfo{author}{\bibfnamefont{J.}~\bibnamefont{Morton}},
  \bibinfo{author}{\bibfnamefont{H.}~\bibnamefont{Wu}},
  \bibinfo{author}{\bibfnamefont{G.}~\bibnamefont{Briggs}},
  \bibinfo{author}{\bibfnamefont{B.}~\bibnamefont{Buckley}},
  \bibinfo{author}{\bibfnamefont{D.}~\bibnamefont{Awschalom}},
  \bibnamefont{et~al.}, \bibinfo{journal}{Phys. Rev. Lett.}
  \textbf{\bibinfo{volume}{105}}, \bibinfo{pages}{140501}
  (\bibinfo{year}{2010}).

\end{thebibliography}
\end{document}